\documentclass[10pt,aps,pra,amsfonts,amssymb,amsmath,
floatfix,nofootinbib,groupedaddress,superscriptaddress,citesort,twocolumn
]{revtex4}%

\usepackage{amsthm}
\usepackage{mathrsfs}
\usepackage{amsfonts}
\usepackage{amstext}
\usepackage{amsmath}
\usepackage{amssymb}
\usepackage{color,xcolor}
\usepackage{float}
\usepackage{bm}
\usepackage{bbm}
\usepackage{graphicx}
\def\qed{\leavevmode\unskip\penalty9999 \hbox{}\nobreak\hfill
	\quad\hbox{\leavevmode  \hbox to.77778em{%
			\hfil\vrule   \vbox to.675em%
			{\hrule width.6em\vfil\hrule}\vrule\hfil}}
	\par\vskip3pt}

\theoremstyle{definition}
\newtheorem{theorem}{Theorem}

\newtheorem{proposition}[theorem]{Proposition}
\newtheorem{example}[theorem]{Example}

\begin{document}
	
	\preprint{APS/123-QED}
	
	\title{Detecting $k$-nonseparability and $k$-partite Entanglement with Generalized Skew Information and Mutually Unbiased Measurements}

	\author{Xiaofei Qi}
\email{xiaofeiqisxu@aliyun.com}
	\affiliation{School of Mathematics and Statistics, Shanxi University,
		Taiyuan 030006, P. R. China}
			\affiliation{Key Laboratory of Complex Systems and Data Science of Ministry of Education,
			Shanxi University, Taiyuan  030006,  Shanxi, China}
			
			\author{Yuyang Pang}
		\email{202212211015@email.sxu.edu.cn}
		\affiliation{School of Mathematics and Statistics, Shanxi University,
			Taiyuan 030006, P. R. China}

	\author{Jinchuan Hou}
\email{jinchuanhou@aliyun.com}
	\affiliation{College of Mathematics, Taiyuan University of
		Technology, Taiyuan 030024, P. R. China}

\begin{abstract}

Multipartite quantum entanglement, as a core quantum resource, is fundamental to the advancement of quantum science and technology.  In multipartite quantum systems, there are two kinds of  quantum entanglement:
$k$-nonseparability and $k$-partite entanglement.  In this paper, we  propose sufficient criteria for detecting \(k\)-nonseparability and \(k\)-partite entanglement by using the generalized Wigner-Yanase skew information and mutually unbiased measurements. Examples are given to demonstrate the detection capability and advantages of these criteria.  As an application, an example  of recognizing the networks by detecting the depth of quantum networks is given.

\end{abstract}


\maketitle

\section{Introduction}

Quantum entanglement plays a crucial role in quantum information science and has important applications
in many fields, such as quantum computing, quantum communication, quantum teleportation, quantum
key distribution and quantum measurements \cite{1,2,3,4,5,6}. By virtue of entanglement, multi-qubits
are allowed to establish intricate nonclassical correlations, which enhance the power of quantum
computing, reinforce  the security of quantum communication, and increase a remarkable
measurement precision.

In quantum entanglement, entanglement detection is one of the main research topics.
For bipartite quantum systems, many criteria for entanglement detection have been proposed
(for example, see \cite{PPT,GHB,CW,H,bloch,GGHE,QH,ZFFW,ZZJW,RL1}  and the references therein). However, in multipartite quantum systems, the task of detecting entanglement grows increasingly complex and challenging with the expansion in the number of parties and the dimension of the systems.
So far, there are two kinds of definitions for multipartite quantum entanglement.
One is defined as ``$k$-nonseparability" in terms of  some subsystems bing entangled while others  not, and another is
defined as ``$k$-partite entanglement" according to the difficulty of preparing entangled states,  i.e., the depth of the entanglement \cite{GOTG,DCT,DC,GYv}.

For the detection of $k$-nonseparability,  some methods are proposed, such as that based on
semidefinite programming \cite{WCXMF}, the entropic uncertainty relations \cite{BG}, quantum Fisher information \cite{HLS,HQGY},
and the generalized Wigner-Yanase skew information \cite{HXGGY}.
For the detection of $k$-partite entanglement, several criteria have also been developed, such as that
based on Wigner-Yanase skew information \cite{C} and  the permutation operators \cite{HGY}.
For other related results, the readers can refer to  \cite{GHL,LGY,HPd,WH,RL,VHE,HQGY2}
and the references therein.
Although some approaches to entanglement detection have been developed from different viewpoints, it remains extremely difficult to identify $k$-nonseparability and $k$-partite entanglement for general mixed states.

In \cite{WF}, the authors gave the concept of mutually unbiased measurements (MUMs) which offers a new
perspective and tools for the flow of information and entanglement detection.
MUMs have significant advantages in quantum information processing, quantum computing and
quantum communication, including enhancing measurement accuracy, reducing errors, and
improving entanglement detection efficiency. Based on MUMs, several criteria
were proposed  for multipartite quantum entanglement detection \cite{LGY,CMF,SC}.

In this paper, we will continue to discuss the topic of detecting $k$-nonseparability and $k$-partite entanglement based on
MUMs and the generalized Wigner-Yanase skew information.
The structure of this paper is as follows. In Section 2, we give some background knowledge on
quantum informations, $k$-separability, and $k$-partite entanglement in multipartite quantum systems.
Section 3 is devoted to derive criteria for detecting $k$-nonseparability and $k$-partite entanglement
in multipartite quantum systems by combining MUMs and the generalized Wigner-Yanase skew information.
In addition, we provide examples to demonstrate the effectiveness of these criteria as well as an application to distinct quantum networks. Section 4 is a conclusion.

\section{Preliminaries}

In this section, we review some important concepts in quantum information, mutually unbiased measurements, $k$-separability and $k$-partite entanglement in multipartite quantum systems.

Throughout this paper, assume that $H_i$ is any complex Hilbert space with $\dim H_i=d_i<\infty$, $i=1,2,\cdots, N$, and let $H=H_1\otimes H_2\otimes \cdots \otimes H_N$.
Denote by $\mathcal B(H)$ the algebra of all bounded linear operators and   ${\mathcal S}(H)=\{ \rho : \rho\geq 0, {\rm Tr}(\rho)=1\}$ is the closed convex set of all quantum states in $\mathcal B(H)$.  $\mathbb{I}\in{\mathcal B}(H)$ and $\mathbb{I}_i\in{\mathcal B}(H_i)$ stand  for the identity operators on $H$ and $H_i$, respectively.

\subsection{$k$-nonseparability and \(k\)-partite entanglement }
For $2\leq k\leq N$,
assume that $\gamma_1,\cdots, \gamma_k$ are non-empty subsets of $\{1, 2,\cdots, N\}$. We say that
$\{\gamma_1,\gamma_2,\cdots,\gamma_k\}$ is   a $k$-partition, denoted by $\gamma=\gamma_1|\gamma_2|\ldots|\gamma_k$,  if it satisfies  the conditions
$\gamma_i \cap \gamma_j=\emptyset $ for  $ i\neq j$
and $\gamma_1\cup \cdots \cup \gamma_k=\{1,2,\cdots,N\}$.

A pure state $|\psi\rangle\langle\psi|\in \mathcal S(H)$ is said to be $k$-separable if there exists a $k$-partition
$\gamma=\gamma_1|\gamma_2|\cdots|\gamma_k$ of
$\{1, 2, \cdots, N\}$  such that
$$|\psi\rangle = \otimes_{i=1}^{k} |\psi_{\gamma_i}\rangle,$$
where $|\psi_i^{\gamma_i}\rangle\langle\psi_i^{\gamma_i}|\in \mathcal S(H_{\gamma_i})$,
$H_{\gamma_i}=\otimes_{j=1}^{k_i}H_{n_{ij}}$, $\gamma_i=\{n_{i1},n_{i2},\cdots,n_{ik_i}\}$,
$n_{i1}<n_{i2}<\cdots<n_{ik_i}$; otherwise, $|\psi\rangle\langle\psi|$ is called $k$-nonseparable.
A mixed state $\rho \in \mathcal S(H)$ is $k$-separable if
$\rho$ can be written as $\rho = \sum_r p_r |\psi_r\rangle\langle\psi_r|$,
where each $|\psi_r\rangle\langle\psi_r|\in{\mathcal S}(H)$ is a $k$-separable pure state and $p_r\geq0$ with $\sum_rp_r=1$;
otherwise, $\rho$ is  $k$-nonseparable.
Particularly,  $\rho$ is   fully separable  if $\rho$ is $N$-separable; and is   genuine entangled
if $\rho$ is 2-nonseparable.

On the other hand, a pure state $|\psi\rangle\langle\psi|\in \mathcal S(H)$ is $k$-producible if it can be represented as
$$
|\psi\rangle = \otimes_{i=1}^{m} |\psi_{\gamma_i}\rangle,
$$
where $\gamma_1|\gamma_2|\cdots|\gamma_m$ is a $m$-partition
of $\{1, 2, \cdots, N\}$, and the number of particles in each subset $\gamma_i$ is not greater than $k$.
For any  mixed state $\rho$, we say that $\rho$ is $k$-producible if it can be represented as a convex combination of \( k \)-producible pure states; otherwise, we say that $\rho$ contains $(k+1)$-partite entanglement, or is $(k+1)$-partite entangled \cite{GOTG}.

Denote by $\mathcal S_k$ the set of all $k$-separable quantum states ($2 \leq k \leq N$) and by $ \mathcal P_k$ the set of all $k$-producible quantum states ($1 \leq k \leq N-1$). By the above definitions, it is easily seen that, if a state is $k$-separable, then it is also $(k-1)$-separable; and if a state is $k$-producible, then it is also $(k+1)$-producible.
That is, $$\mathcal S_N\subset \mathcal S_{N-1}\subset\cdots \subset \mathcal S_2\subset \mathcal S_1; \ \ \mathcal P_N\supset \mathcal P_{N-1}\supset\cdots \mathcal P_2\supset \mathcal P_1,$$
where both $\mathcal S_N$ and $\mathcal P_1$ stand for the set of all fully separable states, while $\mathcal S_1=\mathcal P_N=\mathcal S(H)$.

\subsection{Wigner-Yanase skew information}
In the study of measurement theory, Wigner and Yanase introduced the Wigner-Yanase skew information  from an information-theoretic perspective.
For a quantum state $\rho\in{\mathcal S}(H)$, the Wigner-Yanase skew information of $\rho$ with respect to an observable $A$ is defined by $$I(\rho, A) = -\frac{1}{2} (\text{tr}[\rho^{1/2}, A]^2)$$
with $[A,B]=AB-BA$ as usual \cite{WY}. This quantity $I(\rho, A)$
represents the amount of information about the values of observables that do not commute with \(A\), which is skew-symmetric with respect to $A$.

\subsection{ Quantum Fisher information}

 The quantum Fisher information of a quantum state $\rho\in{\mathcal S}(H)$
with respect to an observable $A\in{\mathcal B}(H)$ is defined as   \cite{HWC}
$$F(\rho, A)=\frac{1}{4}{\rm tr} (\rho L^{2}),$$
 where $L$ is the symmetric logarithmic derivative operator defined via
 $$i(\rho A-A \rho)=\frac{1}{2}(\rho L+L \rho).$$
If \(\rho\) has  the spectral decomposition $\rho=\sum_{l} \lambda_{l}|l\rangle\langle l|$, then $F(\rho,A)$ can be expressed as
 $$F(\rho, A)=\sum_{l, l'} \frac{\left(\lambda_{l}-\lambda_{l'}\right)^{2}}{2\left(\lambda_{l}+\lambda_{l'}\right)}\left|\left< l|A| l'\right>\right|^{2}.$$

\subsection{Generalized Wigner-Yanase skew information }
For any $s\in(-\infty,0)$, define a binary function $f_s$ on $\mathbb R_+\times\mathbb R_+$ ($\mathbb R_+=[0,+\infty)$) as follows:
\[
f_{s}(a, b) = \begin{cases}
\left(\frac{a^{s} + b^{s}}{2}\right)^{1 / s} & \text{if} \ a > 0, b > 0, \\
0 & \text{if} \ a = 0 \ \text{or} \ b = 0,
\end{cases}
\]
which represents the generalized mean of two positive real numbers \( a \) and \( b \) with equal weights.
In addition, define 
\[
f_{0}(a, b) := \lim_{s \to 0} f_{s}(a, b) = \sqrt{ab}
\]
and
\[
f_{-\infty}(a,b) := \lim_{s \to -\infty} f_{s}(a,b) = \min \{a,b\}.
\]
For any quantum state $\rho\in{\mathcal S}(H)$ with  the spectral decomposition
\( \rho = \sum_{l} \lambda_{l}|\psi_{l}\rangle \langle \psi_{l}| \), 
the generalized Wigner-Yanase skew information about an observable \( A \) is given by \cite{YQ}
\[
\begin{aligned}
I^{s}(\rho, A) :=  &\text{tr}\left(\rho A^{2}\right) - \sum_{l, l'} f_{s}\left(\lambda_{l}, \lambda_{l'}\right)\left|\left<\psi_{l}|A|\psi_{l'}\right>\right|^{2} \\
=&  \sum_{l \neq l'} \left[\lambda_{l} - f_{s}\left(\lambda_{l}, \lambda_{l'}\right)\right] \left|\left<\psi_{l}|A|\psi_{l'}\right>\right|^{2}.
\end{aligned}
\]
Particularly, when \( s = -1 \) and \( s = 0 \),  \( I^{-1}(\rho, A) \) and \( I^{0}(\rho, A) \) reduce to the quantum Fisher information and the Wigner-Yanase skew information, respectively.

The authors in \cite{YQ} proved that the generalized Wigner-Yanase skew information has the following properties:

(1) Monotonicity:
$I^s(\rho,A)$ decreases monotonically with respect to \(s\), that is,
$$
I^{0}(\rho, A) \leq I^{-1}(\rho, A) \leq \cdots \leq I^{-\infty}(\rho, A) \leq V(\rho, A),
$$
where $V(\rho, A)= \text{Tr}(\rho A^{2}) - [\text{Tr}(\rho A)]^{2}$ stands for the variance. Moreover, the  equalities hold  if $\rho$ is pure.

(2) Convexity: For any $s$, $I^s$ is convex, that is,
$$
I^{s}\left(\sum_{i} p_{i} \rho_{i}, A\right) \leq \sum_{i} p_{i} I^{s}\left(\rho_{i}, A\right),
$$
where $\rho_i\in{\mathcal S}(H)$, \(p_{i} \geq 0\) and \(\sum_{i} p_{i} = 1\).

(3) Additivity: For any $\rho_{i}\in{\mathcal S}(H_i)$ and any observable $A_i\in\mathcal B(H_i)$, $i=1,2,\cdots,N$, we have
$$
I^{s}\left(\otimes_{i=1}^{N} \rho_{i}, \sum_{i=1}^{N} {\bf A_{i}}\right) = \sum_{i=1}^{N} I^{s}\left(\rho_{i}, A_{i}\right),
$$
where ${\bf A_{i}} := (\otimes_{j=1}^{i-1} \mathbb{I}_{j}) \otimes A_{i} \otimes (\otimes_{j=i+1}^{N} \mathbb{I}_{j})$.

\subsection{Mutually unbiased measurements (MUMs)}
Let $P^{(b)} = \{P_n^{(b)} : n = 1, \cdots, d\} \ (b=1,2)$ be two positive-operator-valued measurements (POVMs) on a $d$-dimensional Hilbert space $H_0$.
$P^{(1)}$ and $P^{(2)}$ are called MUMs if they satisfy the following conditions:
$$
\begin{cases}
\mathrm{tr}(P_n^{(b)}) = 1, \\
\mathrm{tr}(P_n^{(b)} P_{n'}^{(b')}) =& \delta_{n,n'} \delta_{b,b'} \kappa \\&+ (1-\delta_{n,n'}) \delta_{b,b'} \frac{1-\kappa}{d-1} + (1-\delta_{b,b'}) \frac{1}{d},
\end{cases}
$$
where $b, b' \in \{1, 2\},\ n, n' \in \{1, \cdots, d\}$, and the parameter \( \kappa \) satisfies \( \frac{1}{d} < \kappa \leq 1 \). When \( \kappa = 1 \), it reduces to mutually unbiased bases (MUBs) \cite{WF}.

Kalev and Gour  in \cite{KG} proposed a construction method for MUMs in any dimensional quantum system.
Take an orthonormal basis $\{F_k\}_{k=1}^{d^{2} - 1}$ for $B_0(H_0)$, consisting of Hermitian and traceless operators acting on $H_0$  and  with inner product $(A,B)={\rm Tr} (B^\dag A)$.
Arrange these basis elements on a grid of $d-1$ columns and $d+1$ rows: the first row consists of
$F_{1}, F_{2}, \ldots, F_{d-1}$,
the second row consists of
$F_{d}, F_{d+1}, \ldots, F_{2(d-1)}$,
and so on.
Denote the element in the $n$-th column of the $b$-th row by
$F_{n,b}$, with $n = 1,2,\ldots,d-1$ and $b = 1,2,\ldots,d+1$.
Define $d(d+1)$ operators $F_n^{(b)}$ by
\begin{equation}
F_n^{(b)} = \left\{
\begin{array}{ll}
F^{(b)} - (d + \sqrt{d}) F_{n, b} & \text{for} \ n=1, \cdots, d-1, \\
(1 + \sqrt{d})F^{(b)} & \text{for} \ n=d,
\end{array}
\right.
\end{equation}
where $F^{(b)}=\sum_{n=1}^{d-1} F_{n, b}$, \(b = 1, 2,\cdots, d+1\). 
It is a direct calculation that
\begin{equation}
\sum_{n=1}^{d}F_{n}^{(b)}=0,\ \ \forall b\in\{1, 2,\cdots, d+1\}.
\end{equation}
It was proved in \cite{KG} that there exists a parameter $t$, called the free parameter, such that the   operators $P_n^{(b)}$  given by
\begin{equation}
P_n^{(b)} = \frac{1}{d}\mathbb{I} + t F_n^{(b)}
\end{equation}
are positive for all $n,b$. Put
\begin{equation}
\kappa=\frac{1}{d}+t^2(1+\sqrt{d})^2(d-1),
\end{equation}
then $P^{(b)}=\{P_n^{(b)}\}_{n=1}^{d}$ ($b=1,2,\cdots,d+1$) form a complete set of \(d+1\) MUMs, where $b$ labels the measurement and $n$ labels the outcome. Moreover, any complete set of MUMs has this form and satisfies the following equation:
\begin{equation}
\sum_{b=1}^{d+1} \sum_{n=1}^{d} (P_n^{(b)})^2 = (d+1)\kappa \mathbb{I}.
\end{equation}

 It was also shown in \cite{LL} that
$$\left\{\frac{\mathbb{I}}{\sqrt{d}}, F_{n,b} \colon n=1,\dots,d-1;\, b=1,\dots,d+1\right\}$$ forms  a set of local orthogonal observables (LOOs) on $H_0$.

\section{\(k\)-separability and \(k\)-partite entanglement detection criteria}

In this section, we will establish some entanglement detection criteria for \(k\)-separability and \(k\)-partite entanglement in multipartite quantum systems.

For convenience, assume
$H_i = H_0$ with $\dim H_0 = d$,  $i = 1,2,\dots,N$.
Take the same   $d+1$ MUMs on each subsystem $H_i$:
$$P^{(i,b)} = \{ P_{i,n}^{(b)}\}_{n=1}^d\ \ (b = 1, 2, \cdots, d+1),$$
where \(i\) is used to label the subsystem on which the operators act,  \(i = 1, 2, \cdots, N\). 
For any subset \(\Gamma \subset \{1, 2, \cdots, N\}\) with $\sharp(\Gamma)$  the number of the elements in  $\Gamma$, 
write
\[
{\bf P}_{\Gamma,n}^{(b)} = \sum_{i \in \Gamma} {\bf P}_{i,n}^{(b)} \]
with\[{\bf P}_{i,n}^{(b)} = (\otimes_{j=1}^{i-1} \mathbb{I}_{j}) \otimes P_{i,n}^{(b)} \otimes (\otimes_{j=i+1}^{\sharp(\Gamma)} \mathbb{I}_{j});
\]
and in particular, when \(\Gamma = \{1, 2, \cdots, N\}\),  write
$${\bf P}_{N,n}^{(b)} = \sum_{i=1}^N {\bf P}_{i,n}^{(b)}.$$

For ${\bf P}_{\Gamma,n}^{(b)}$, we have the following result.

\begin{proposition}
For any \( p \neq q \in \Gamma \),  we have
\[
\sum_{b=1}^{d+1}\sum_{n=1}^d {\bf P}_{p,n}^{(b)} {\bf P}_{q,n}^{(b)} \leq (1 + \kappa)(\otimes_{i\in\Gamma} \mathbb{I}_{i}).
\]
\end{proposition}

\begin{widetext}
{\bf Proof.} Without loss of generality, assume $p<q$.
By the definition, one has
\[
{\bf P}_{p,n}^{(b)} {\bf P}_{q,n}^{(b)} = (\otimes_{j=1}^{p-1} \mathbb{I}_{j}) \otimes P_{p,n}^{(b)} \otimes (\otimes_{j=p+1}^{q-1} \mathbb{I}_{j}) \otimes P_{q,n}^{(b)} \otimes (\otimes_{j=q+1}^{\sharp(\Gamma)} \mathbb{I}_{j}).
\]
It follows from Eqs.(1)-(4) that
$$\begin{aligned}
&\sum_{b=1}^{d+1}\sum_{n=1}^d{\bf P}_{p,n}^{(b)}{\bf P}_{q,n}^{(b)}\\
=&\sum_{b=1}^{d+1}\sum_{n=1}^d (\otimes_{j=1}^{p-1} \mathbb{I}_{j}) \otimes (\frac{1}{d}\mathbb{I}_p+tF_{n}^{(b)}) \otimes (\otimes_{j=p+1}^{q-1} \mathbb{I}_{j}) \otimes (\frac{1}{d}\mathbb{I}_q+tF_{n}^{(b)}) \otimes (\otimes_{j=q+1}^{\sharp(\Gamma)} \mathbb{I}_{j})\\
=&\sum_{b=1}^{d+1}\sum_{n=1}^d\big[\frac{(\otimes_{i\in\Gamma} \mathbb{I}_{i})}{d^2}+\frac{t}{d}((\otimes_{j=1}^{p-1} \mathbb{I}_{j})\otimes F_n^{(b)}\otimes(\otimes_{j=p+1}^{\sharp(\Gamma)} \mathbb{I}_{j})+(\otimes_{j=1}^{q-1} \mathbb{I}_{j})\otimes F_n^{(b)}\otimes(\otimes_{j=q+1}^{\sharp(\Gamma)} \mathbb{I}_{j}))\\
&+t^2(\otimes_{j=1}^{p-1} \mathbb{I}_{j}) \otimes F_n^{(b)} \otimes (\otimes_{j=p+1}^{q-1} \mathbb{I}_{j}) \otimes F_n^{(b)} \otimes (\otimes_{j=q+1}^{\sharp(\Gamma)} \mathbb{I}_{j})\big]\\
=&\frac{d(d+1)}{d^2}(\otimes_{i\in\Gamma} \mathbb{I}_{i})+t^2\sum_{b=1}^{d+1}\sum_{n=1}^d(\otimes_{j=1}^{p-1} \mathbb{I}_{j}) \otimes F_n^{(b)} \otimes (\otimes_{j=p+1}^{q-1} \mathbb{I}_{j}) \otimes F_n^{(b)} \otimes (\otimes_{j=q+1}^{\sharp(\Gamma)} \mathbb{I}_{j})\\
=&\frac{d+1}{d}(\otimes_{i\in\Gamma} \mathbb{I}_{i})+t^2(d+\sqrt{d})^2\sum_{b=1}^{d+1}\sum_{n=1}^{d-1}(\otimes_{j=1}^{p-1} \mathbb{I}_{j}) \otimes F_{n,b} \otimes (\otimes_{j=p+1}^{q-1} \mathbb{I}_{j}) \otimes F_{n,b} \otimes (\otimes_{j=q+1}^{\sharp(\Gamma)} \mathbb{I}_{j})\\
\leq&\frac{d+1}{d}(\otimes_{i\in\Gamma} \mathbb{I}_{i})+t^2(d+\sqrt{d})^2(1-\frac{1}{d})(\otimes_{i\in\Gamma} \mathbb{I}_{i})\\
=&(1+\kappa)(\otimes_{i\in\Gamma} \mathbb{I}_{i}).
\end{aligned}$$
Here, the first equality uses Eq.(3), the third equality uses Eq.(2), the fourth equality uses Eq.(1) and the last equality uses Eq.(4).
\hfill$\Box$
\end{widetext}

\begin{proposition}
Assume that $\Gamma\subset\{1,2,\cdots,N\}$ is any subset.  For any $\rho\in{\mathcal S}(H_1\otimes H_2\otimes\cdots\otimes H_N)$,  denoting by  \(\rho_{\Gamma}\) the reduced state of \(\rho\) on the subsystem \(\otimes_{i \in \Gamma} H_i\), we have
$$
\sum_{b=1}^{d+1}\sum_{n=1}^d I^s(\rho_{\Gamma},{\bf P}_{\Gamma,n}^{(b)})\leq \sharp(\Gamma)^2(\kappa-\frac{1}{d})+\sharp(\Gamma)(d\kappa-1).
$$
\end{proposition}

{\bf Proof.}
For any subset $\Gamma\subset\{1,2,\cdots,N\}$, by Proposition 3.1 and Eq.(5), we have
\[
\begin{aligned}
& \sum_{b=1}^{d+1}\sum_{n=1}^d\big({\bf P}_{\Gamma,n}^{(b)}\big)^2\\
= &\sum_{b=1}^{d+1}\sum_{n=1}^d\big(\sum_{i\in\Gamma}{\bf P}_{i,n}^{(b)}\big)^2\\
=& \sum_{b=1}^{d+1}\sum_{n=1}^d \big(\sum_{i\in\Gamma}({\bf P}_{i,n}^{(b)})^2+\sum_{p\not= q\in\Gamma}{\bf P}_{p,n}^{(b)}{\bf P}_{q,n}^{(b)}\big)\\
= &\sum_{i\in\Gamma}\sum_{b=1}^{d+1}\sum_{n=1}^d((\otimes_{j=1}^{i-1} \mathbb{I}_{j})\otimes P_{i,n}^{(b)}\otimes (\otimes_{j=i+1}^{\sharp(\Gamma)} \mathbb{I}_{j}))^2\\&+\sum_{p\neq q\in\Gamma}\sum_{b=1}^{d+1}\sum_{n=1}^d{\bf P}_{p,n}^{(b)}{\bf P}_{q,n}^{(b)}\\
\leq & \sharp(\Gamma)(d+1)\kappa (\otimes_{i\in\Gamma} \mathbb{I}_{i})+(\sharp(\Gamma)^2-\sharp(\Gamma))(1+\kappa)(\otimes_{i\in\Gamma} \mathbb{I}_{i})\\
= & [\sharp(\Gamma)^2(1+\kappa)+\sharp(\Gamma)(d\kappa-1)](\otimes_{i\in\Gamma} \mathbb{I}_{i}).
\end{aligned}
\]
On the other hand, by Eq.(2),
one has
\[
\begin{aligned}
&\sum_{b=1}^{d+1}\sum_{n=1}^d \big(\mathrm{tr}\rho_{\Gamma}{\bf P}_{\Gamma,n}^{(b)}\big)^2\\
=&\sum_{b=1}^{d+1}\sum_{n=1}^d[\mathrm{tr}\rho_{\Gamma}\sum_{i\in\Gamma}{\bf P}_{i,n}^{(b)}]^2
=\sum_{b=1}^{d+1}\sum_{n=1}^d[\sum_{i\in\Gamma}\mathrm{tr}\rho_{\Gamma}{\bf P}_{i,n}^{(b)}]^2\\
=&\sum_{b=1}^{d+1}\sum_{n=1}^d[\sum_{i\in\Gamma}\mathrm{tr}\rho_{\Gamma}(\frac{\otimes_{i\in\Gamma} \mathbb{I}_{i}}{d}+t{\bf F}_{i,n}^{(b)})]^2\\
=&\sum_{b=1}^{d+1}\sum_{n=1}^d(\frac{\sharp(\Gamma)}{d}+t\sum_{i\in\Gamma}\mathrm{tr}\rho_{\Gamma}{\bf F}_{i,n}^{(b)})^2\\
=&\sum_{b=1}^{d+1}\sum_{n=1}^d\left(\frac{\sharp(\Gamma)}{d}\right)^2
+t^2\sum_{b=1}^{d+1}\sum_{n=1}^d\left(\sum_{i\in\Gamma}\mathrm{tr}\left(\rho_{\Gamma}{\bf F}_{i,n}^{(b)}\right)\right)^2\\
\geq & \sum_{b=1}^{d+1}\sum_{n=1}^d(\frac{\sharp(\Gamma)}{d})^2= \sharp(\Gamma)^2(1+\frac{1}{d}),
\end{aligned}
\]
where ${\bf F}_{i,n}^{(b)} = (\otimes_{j=1}^{i-1} \mathbb{I}_{j}) \otimes F_{i,n}^{(b)} \otimes (\otimes_{j=i+1}^{\sharp(\Gamma)} \mathbb{I}_{j})$.
Combining the above two inequalities gives
$$
\begin{aligned}
&\sum_{b=1}^{d+1}\sum_{n=1}^d I^s(\rho_{\Gamma},{\bf P}_{\Gamma,n}^{(b)})\\
\leq & \sum_{b=1}^{d+1}\sum_{n=1}^d \Big[ \mathrm{tr} \, \rho_{\Gamma} ({\bf P}_{\Gamma,n}^{(b)})^2 - (\mathrm{tr} \, \rho_{\Gamma} {\bf P}_{\Gamma,n}^{(b)})^2 \Big] \\
\leq & [\sharp(\Gamma)^2(1+\kappa)+\sharp({\Gamma})(d\kappa-1)]-\sharp(\Gamma)^2(1+\frac{1}{d})\\
=& \sharp(\Gamma)^2(\kappa-\frac{1}{d})+\sharp(\Gamma)(d\kappa-1).
\end{aligned}
$$ \hfill$\Box$

Now, we can give a \(k\)-nonseparability  criterion and a \(k\)-partite entanglement criterion
by generalized Wigner-Yanase skew informations and MUMs.

\begin{theorem}
If $\rho$ is $k$-separable, then the following inequality holds:
$$
\begin{aligned}
& \sup_{s\in[-\infty,0]}\sum_{b=1}^{d+1}\sum_{n=1}^d I^s(\rho, {\bf P}_{N,n}^{(b)})\\
 =&\sum_{b=1}^{d+1} \sum_{n=1}^{d} I^{-\infty}\left(\rho, \mathbf{P}_{N,n}^{(b)}\right)\\
\leq &N(d\kappa - 1) + \left[(N-k+1)^2 + k-1\right](\kappa - \frac{1}{d}).
\end{aligned}$$
\end{theorem}

{\bf Proof.}
Define the generalized skew information function:
\[
g_s(\rho) = \sum_{b=1}^{d+1} \sum_{n=1}^{d} I^s\left(\rho, \mathbf{P}_{N,n}^{(b)}\right), \quad s \in [-\infty, 0].
\]

Since $\rho$  is \(k\)-separable, $\rho$ can be written as a convex combination of $k$-separable pure states, that is, $\rho =\sum_rp_r\rho_r$, where $\rho^{(r)}$ has the form $$\rho_r=\rho^{(r)}_{\Gamma_1}\otimes\rho^{(r)}_{\Gamma_2}\otimes\cdots\otimes\rho^{(r)}_{\Gamma_k}$$
for some $k$-partition $ \Gamma_1|\Gamma_2|\ldots|\Gamma_k$ of $\{1,2,\cdots,N\}$.
By Proposition 2, and by the convexity and additivity of the generalized Wigner-Yanase skew information, we have
$$
\begin{aligned}
g_s(\rho)=& \sum_{b=1}^{d+1}\sum_{n=1}^d I^s(\rho, {\bf P}_{N,n}^{(b)})\\
\leq &\sum_{b=1}^{d+1}\sum_{n=1}^d\sum_rp_rI^s(\rho_r, {\bf P}_{N,n}^{(b)})\\ =& \sum_{b=1}^{d+1}\sum_{n=1}^d\sum_r p_r\sum_{t=1}^kI^s(\rho_{\Gamma_t}^{(r)},{\bf P}_{\Gamma_t,n}^{(b)})\\
\leq& \sum_r p_r\sum_{i=1}^k[\sharp(\Gamma_i)^2(\kappa-\frac{1}{d})+\sharp(\Gamma_i)(d\kappa-1)]\\ = &(\kappa-\frac{1}{d})\sum_{i=1}^k\sharp(\Gamma_i)^2+N(d\kappa-1).
\end{aligned}
$$

The next goal is to maximize $\sum_{i=1}^k\sharp(\Gamma_i)^2$.
To do this, write
$$x = (\sharp(\Gamma_{i_1}), \sharp(\Gamma_{i_2}), \dots, \sharp(\Gamma_{i_k}))^t \in \mathbb{R}^k,$$
where
$\sharp(\Gamma_{i_j}) \in \mathbb{Z}^+$ (the set of positive integers), $\sharp(\Gamma_{i_1}) \geq \sharp(\Gamma_{i_2}) \geq \dots \geq \sharp(\Gamma_{i_k})$ with $\sum_{j=1}^k \sharp(\Gamma_{i_j}) = N$, $i_j\not=i_l\in\{1,2,\cdots,k\}$.
Denote by $X\subset \mathbb{R}^k$ the set consisting of all the above vectors $x$. Write
$$y_0 = (N - k + 1, 1, \cdots, 1)^t \in \mathbb{R}^k.$$
It is obvious that $y_0\in X$.
Recall that, for any nonnegative vectors $x=(x_1,x_2,\cdots,x_k)^t$ and $y=(y_1,y_2,\cdots,y_k)^t\in{\mathbb R}^k$ with coordinate descending order arrangement,
$x \prec y$ means that
$\sum_{i=1}^mx_i \leq \sum_{i=1}^my_i$ holds for all $m=1,2,\cdots,k-1$ and $\sum_{i=1}^kx_i =\sum_{i=1}^ky_i$.
Then we have $x \prec y_0$ for all $x\in X$; otherwise,
if there exists some $m'$ such that
$\sum_{i=1}^{m'} x_i > \sum_{i=1}^{m'} y_i =N - k + m',$
then there must be some $x_0=0$, a contradiction.
Note that the function $x\mapsto\sum_{i=1}^kx_i^2 $ is Schur convex. It follows that
$\sum_{i=1}^kx_i^2 \leq (N - k + 1)^2 + (k-1)$. This implies that  $\sum_{i=1}^k\sharp(\Gamma_i)^2$
attains the maximum at the partition $\{\Gamma_1,\cdots,\Gamma_k\}$ with
$\sharp(\Gamma_1) = N - k + 1,\ \sharp(\Gamma_i)=1$ for all $i=2,3,\cdots,k$.

Hence, for any $s \in [-\infty, 0]$, one obtains
\[
g_s(\rho) \leq N(d\kappa-1) + \left[(N-k+1)^2 + k-1\right]\left(\kappa - \frac{1}{d}\right).
\]

Now,  by the monotonicity of the generalized Wigner-Yanase skew information, we achieve
\[\begin{array}{rl}&\sup_{s \in [-\infty, 0]} g_s(\rho) = g_{-\infty}(\rho)\\ = &\sum_{b=1}^{d+1} \sum_{n=1}^{d} I^{-\infty}\left(\rho, \mathbf{P}_{N,n}^{(b)}\right)\\
	\leq&N(d\kappa-1) + \left[(N-k+1)^2 + k-1\right]\left(\kappa - \frac{1}{d}\right),
\end{array}\]
as desired. \hfill$\Box$

\begin{theorem}
If  $\rho$ is $k$-producible, then the following inequality holds:
\begin{widetext}$$
\begin{array}{rl} & \sup_{s\in[-\infty,0]}\sum_{b=1}^{d+1}\sum_{n=1}^d I^s(\rho, {\bf P}_{N,n}^{(b)})
= \sum_{b=1}^{d+1} \sum_{n=1}^{d} I^{-\infty}(\rho, {\bf P}_{N,n}^{(b)}) \\ \leq &
\begin{cases}
Nk(\kappa - \frac{1}{d}) + N(d\kappa-1) & {\rm if}\ \ N=pk; \\
(N^2+p^2k^2+pk^2 -2pkN) (\kappa - \frac{1}{d}) + N(d\kappa-1) & {\rm if}\ \ 0<N-pk<k,
\end{cases}
\end{array}$$
where $p=[\frac{N}{k}]$.\end{widetext}\end{theorem}

{\bf Proof.}
Since $\rho$ is \(k\)-producible,  $\rho$ can be written as \(\rho = \sum_r p_r \rho_r\), where
$$\rho_r = \rho^{(r)}_{\Gamma_1} \otimes \rho^{(r)}_{\Gamma_2} \otimes \cdots \otimes \rho^{(r)}_{\Gamma_{l_r}}$$
for some $l_r$-partition $\Gamma_1|\Gamma_2|\ldots|\Gamma_{l_r}$ with $\sharp(\Gamma_i) \leq k$  for every $i=1,\cdots, l_r\leq N$. Let $\sharp(\Gamma_i)=0$ for $i=l_r+1,\ldots, N$. Then, by Proposition 3.2, we have
$$
\begin{aligned}
&\sum_{b=1}^{d+1}\sum_{n=1}^d I^s(\rho, {\bf P}_{N,n}^{(b)})\\
\leq & \sum_{b=1}^{d+1}\sum_{n=1}^d\sum_rp_rI^s(\rho_r, {\bf P}_{N,n}^{(b)})\\
=&\sum_{b=1}^{d+1}\sum_{n=1}^d\sum_r p_r\sum_{t=1}^{l_r}I^s(\rho_{\Gamma_t}^{(r)},{\bf P}_{\Gamma_t,n}^{(b)})\\
\leq & \sum_r p_r\sum_{i=1}^{l_r}[\sharp(\Gamma_i)^2(\kappa-\frac{1}{d})+\sharp(\Gamma_i)(d\kappa-1)]\\
=&(\kappa-\frac{1}{d})\sum_{i=1}^{N}\sharp(\Gamma_i)^2+N(d\kappa-1).
\end{aligned}
$$
Write
$$x = (\sharp(\Gamma_{i_1}), \sharp(\Gamma_{i_2}), \dots, \sharp(\Gamma_{i_{l_r}}),0,\ldots,0)^t \in \mathbb{R}^{N},$$
where
$\sharp(\Gamma_{i_j}) \in \mathbb{Z}^+$ with $\sharp(\Gamma_{i_j})\leq k$, $\sharp(\Gamma_{i_1}) \geq \sharp(\Gamma_{i_2}) \geq \dots \geq \sharp(\Gamma_{i_{l_r}})$ with $\sum_{j=1}^{l_r} \sharp(\Gamma_{i_j}) = N$, $i_j\not=i_p\in\{1,2,\cdots,l_r\}$.
Denote by $Y\subset \mathbb{R}^N$ the set consisting of all the above vectors $x$. Write
$$y_0 =(\underbrace{k, k, \dots, k}_{p}, N-pk, 0, \dots, 0)^t \in \mathbb{R}^N,$$
where $p=[\frac{N}{k}]$ the integer part of $\frac{N}{k}$.
It is easily seen  that $y_0\in Y$ and  $x \prec y_0$ for all $x\in Y$. By the Schur convexity of
the function $x\mapsto\sum_{i=1}^Nx_i^2 $ again, we see that
$\sum_{i=1}^Nt_i^2$
attains the maximum at $y_0$. Consequently, $\sum_{i=1}^N\sharp{(\Gamma_i)}^2$  achieves the maximum at  the $(p+1)$-partition $\{\Gamma_1,\cdots,\Gamma_{l}\}$, where $l=p$ if $N=pk$ and $l=p+1$ if $N\not=pk$, with
$$
\sharp(\Gamma_1) = \sharp(\Gamma_2) = \dots = \sharp(\Gamma_p) = k,
\quad \sharp(\Gamma_{p+1}) =N-pk.$$
Thus, if $N =pk$, then
$$\begin{array}{rl}& \sum_{b=1}^{d+1}\sum_{n=1}^d I^s(\rho, {\bf P}_{N,n}^{(b)})\\
\leq & pk^2(\kappa-\frac{1}{d})+N(d\kappa-1)=Nk(\kappa - \frac{1}{d}) + N(d\kappa-1);
\end{array}$$
and if $N=pk+m$ with $0<m<k$, then
$$
\begin{array}{rl}
&\sum_{b=1}^{d+1}\sum_{n=1}^d I^s(\rho, {\bf P}_{N,n}^{(b)})\\
 \leq &  pk^2(\kappa-\frac{1}{d})+(N-pk)^2(\kappa-\frac{1}{d})+N(d\kappa-1)\\
= & (N^2+p^2k^2+pk^2 -2pkN)(\kappa - \frac{1}{d}) + N(d\kappa-1).
\end{array}
$$
Thus, the monotonicity of generalized Wigner-Yanase skew information ensures that the tightest bounds are achieved in the limiting case $s= -\infty$. This completes the proof of the theorem.
\hfill$\Box$

Consequently, any violation of the inequalities in Theorems 3-4 indicate $k$-nonseparability and $(k+1)$-partite entanglement, respectively.

We remark here that, although the supremum over $s \in [-\infty,0]$ in Theorems 3-4 is theoretically difficult to evaluate due to the nonlinear dependence of $\sum_{b=1}^{d+1}\sum_{n=1}^d I^s(\rho,{\bf P}_{N,n}^{(b)})$ on $s$,
in practice it can be well approximated by considering extreme values such as $s=-\infty$ or by sampling $s$ at a finite set of points.
This allows effective detection of $k$-nonseparability while retaining the practical advantages of the criterion.

To verify the validity of these criteria, we provide several examples below.

\begin{example}
Consider the 6-partite quantum state on the Hilbert space $H = (\mathbb{C}^2)^{\otimes 6}$:
$$
\rho(p)=p|\psi\rangle\langle\psi|+\frac{1-p}{64}\mathbb{I},
$$
where $|\psi\rangle=\frac{1}{\sqrt{6}}(|000001\rangle+|000010\rangle+|000100\rangle+|001000\rangle+|010000\rangle+|100000\rangle)$ and  $0 \leq p \leq 1$.
\end{example}

In this case, we illustrate the situation by taking $s = -1, 0, -\infty$, respectively.
According to Theorem 3 and Theorem 4, we can obtain the following conclusions:
when $p_k < p \leq 1$, $\rho(p)$ is $k$-nonseparable; when $\tilde{p_k} < p \leq 1$, $\rho(p)$ has $(k + 1)$-partite entanglement.
In \cite{HQGY}, the authors showed that $\rho(p)$ is $k$-nonseparable if $p_k'<p\leq 1$, and $\rho(p)$ contains $(k+1)$-partite entanglement if
$\tilde{p_k'}<p\leq 1$.   See the following tables.

\begin{table}[H]
    \centering
    \setlength{\tabcolsep}{3pt}
    \begin{tabular}{l|ccccc}
        \hline
        $k$ & 2 & 3 & 4 & 5 & 6  \\
        \hline
        $p_k$  & 0.6523  & 0.5211 & 0.4225 & 0.3567 & 0.3237  \\
        $p_k'$ & $\backslash$  & 0.5816 & 0.4433 & 0.3443 & 0.2649  \\
        \hline
    \end{tabular}
    \caption{\small Range for detecting $k$-nonseparability for $s=-1$.}
    \label{tab:k-inseparable}
\end{table}

\begin{table}[H]
    \centering
    \setlength{\tabcolsep}{3pt}
    \begin{tabular}{l|ccccc}
        \hline
        $k$ & 1 & 2 & 3 & 4 & 5 \\
        \hline
        $\tilde{p_k}$ & 0.3237  & 0.4225 & 0.5211 & 0.5539 & 0.6523 \\
        $\tilde{p_k'}$ & 0.2649  & 0.5026 & 0.6210 & 0.6605 & 0.7591 \\
        \hline
    \end{tabular}
    \caption{\small Range for detecting $(k+1)$-partite entanglement for $s=-1$.}
    \label{tab:k-entangled}
\end{table}

\begin{table}[H]
    \centering
    \setlength{\tabcolsep}{3pt}
    \begin{tabular}{l|ccccc}
        \hline
        $k$ & 2 & 3 & 4 & 5 & 6  \\
        \hline
        $p_k$  & 0.9943  & 0.9927 & 0.9913 & 0.9903 & 0.9898  \\
        $p_k'$ & $\backslash$  & 0.9990 & 0.9978 & 0.9961 & 0.9939  \\
        \hline
    \end{tabular}
    \caption{\small Range for identifying $k$-nonseparability for $s=0$.}
    \label{tab:k-inseparable}
\end{table}

\begin{table}[H]
    \centering
    \setlength{\tabcolsep}{3pt}
    \begin{tabular}{l|ccccc}
        \hline
        $k$ & 1 & 2 & 3 & 4 & 5 \\
        \hline
        $\tilde{p_k}$  & 0.9898  & 0.9913 & 0.9927 & 0.9931 & 0.9943 \\
        $\tilde{p_k'}$ & 0.9939  & 0.9961 & 0.9990 & $\backslash$ & $\backslash$ \\
        \hline
    \end{tabular}
    \caption{\small Range for identifying $(k+1)$-partite entanglement for $s=0$.}
    \label{tab:k-entangled}
\end{table}

\begin{table}[H]
    \centering
    \setlength{\tabcolsep}{3pt}
    \begin{tabular}{l|ccccc}
        \hline
        $k$ & 2 & 3 & 4 & 5 & 6  \\
        \hline
        $p_k$  & 0.3958  & 0.3125 & 0.25 & 0.2083 & 0.1875  \\
        $p_k'$ & $\backslash$  & 0.75 & 0.625 & 0.5 & 0.375  \\
        \hline
    \end{tabular}
    \caption{\small Range for identifying $k$-nonseparability for $s=-\infty$.}
    \label{tab:k-inseparable}
\end{table}

\begin{table}[H]
    \centering
    \setlength{\tabcolsep}{3pt}
    \begin{tabular}{l|ccccc}
        \hline
        $k$ & 1 & 2 & 3 & 4 & 5 \\
        \hline
        $\tilde{p_k}$  & 0.1875  & 0.25 & 0.3125 & 0.3333 & 0.3958 \\
        $\tilde{p_k'}$ & 0.375  & 0.5 & 0.75 & $\backslash$ & $\backslash$ \\
        \hline
    \end{tabular}
    \caption{\small Range for identifying $(k+1)$-partite entanglement for $s=-\infty$.}
    \label{tab:k-entangled}
\end{table}

As clearly demonstrated in Tables I-VI, the case $s = -\infty$ exhibits the widest detection range, consistently outperforming both $s=-1$ and $s=0$ in identifying entangled states across all tested scenarios. This observation further corroborates our theoretical finding that the optimal detection capability is achieved precisely at the limiting case $s= -\infty$, where the generalized Wigner-Yanase skew information attains its maximal sensitivity. The enhanced performance at $s=-\infty$ can be attributed to the complete utilization of all high-order quantum coherences in the state detection.

In addition, it can be seen that, Theorem 3 can detect more \(k\)-nonseparable quantum states; and, except for $k=1$,  Theorem 4 can detect more   \((k + 1)\)-partite entangled quantum states, which indicate that our criteria have better effectiveness.\hfill$\Box$

\begin{example}
Consider an 11-partite quantum state on the Hilbert space $H = (\mathbb{C}^2)^{\otimes 11}$:
$$
\rho = p|\psi\rangle\langle\psi|+\frac{1 - p}{2^{11}}\mathbb{I},
$$
where
$|\psi\rangle=\frac{1}{\sqrt{2}}(|0\rangle^{\otimes 11}+|1\rangle^{\otimes 11})$ and  $0 \leq p \leq 1$.
\end{example}

In this example, we take $s = -\infty $ for an illustration.
According to Theorem 3 and Theorem 4, we can conclude that $\rho(p)$ is $k$-nonseparable
if $p_k < p \leq 1$ and exhibits $(k + 1)$-partite entanglement
if $\tilde{p_k} < p \leq 1$.
Note that, $p_k'$ is the result obtained using the method in \cite{HLS}, and $\tilde{p_k'}$ is obtained in \cite{HQGY2}, as shown in the Tables VII-VIII.

\begin{widetext}
\begin{table}[ht]
 \begin{minipage}{\textwidth}
    \centering
    \setlength{\tabcolsep}{10pt} 
    \renewcommand{\arraystretch}{1.5} 
    \begin{tabular}{l|cccccccccc}
        \hline
        \(k\) & 2 & 3 & 4 & 5 & 6 & 7 & 8 & 9 & 10 & 11 \\
        \hline
        \(p_k\)  & 0.4300  & 0.3671 & 0.3111 & 0.2622 & 0.2202 & 0.1853  & 0.1573 & 0.1363 & 0.1223 & 0.1153 \\
        \(p_k'\) & 0.8532  & 0.7205 & 0.6017 & 0.4969 & 0.4061 & 0.3293  & 0.2664 & 0.2175 & 0.1826 & 0.1546 \\
        \hline
    \end{tabular}
    \caption{Range for identifying $k$-nonseparability.}
    \label{tab:k-inseparable}
 \end{minipage}
\end{table}

\begin{table}[ht]
    \centering
    \setlength{\tabcolsep}{10pt} 
    \renewcommand{\arraystretch}{1.5} 
    \begin{tabular}{l|cccccccccc}
        \hline
         \(k\) & 1 & 2 & 3 & 4 & 5 & 6 & 7 & 8 & 9 & 10 \\
            \hline
            $\tilde{p_k}$  & 0.1153  & 0.1503 & 0.1853 & 0.2202 & 0.2552 & 0.2902  & 0.3041 & 0.3321 & 0.3741 & 0.4300 \\
            $\tilde{p_k'}$ & 0.0009  & 0.0312 & 0.1248 & 0.2498 & 0.2498 & 0.4997  & 0.4997 & 0.4997 & 0.4997 & 0.4997 \\
            \hline
    \end{tabular}
    \caption{Range for identifying $(k+1)$-partite entanglement.}
    \label{tab:k-entanglement}
\end{table}
\end{widetext}

From the tables, it can be seen that, for $2 \leq k \leq 11$,  Theorem 3 can identify more $k$-nonseparable quantum states for all $k$ compared to that in \cite{HLS}, which reveals that our criterion has better effectiveness.  For $k = 4, 6, 7, 8, 9, 10$, utilizing  Theorem 4 can detect more $(k + 1)$-partite entangled quantum states compared to that  in \cite{HQGY2}, which   demonstrates that our criterion performs better for biger $k$s.\hfill$\Box$

Finally, we give an example of applying Theorem 4 to distinct the networks by detecting the depth of quantum networks.

A quantum network is a system that connects quantum devices to share quantum information, enabling tasks like secure communication and  distributed computing  using quantum properties  \cite{LMX,KHJ}. Generally,  a quantum network consists of  $n$  nodes (subsystems, denote by $A_j$, $j=1,2,\cdots,n$) and  $r$  quantum resources $\rho_i$ shared by some of the nodes.
Let $L_i \subseteq  \{1,2,\cdots,n\}$ be the subset of  index of the nodes that $\{A_j: j\in L_i\}$ is exact the subset of all nodes that share the $i$-th resource $\rho_i$. Denote by $\rho_i=\rho_{L_i}$,  $i=1,2,\cdots,r$. Then the corresponding
  quantum network state has the form:
$$\rho=\rho_{L_1}\otimes \rho_{L_2}\otimes\cdots\otimes\rho_{L_r}.$$
Accordingly, for every $j=1,2,\ldots, n$, the node $A_j$ may share several $\rho_{L_i}$ with other nodes, that is $j\in L_i$. Let $F_j=\{i_j: A_j\in L_i\}\subseteq\{1,2,\ldots,r\}$. Then the subsystem hold by  $A_j$ is a $\sharp F_j$-partite composite system  $H_{A_j}=\bigotimes_{i_j\in F_j} H_{A_{i_j}}$.
 Every quantum network state determined by this network has the form:
\begin{equation}
\rho=\sum_{\lambda\in\Lambda}p_\Lambda(\lambda)(\Omega_1^\lambda\otimes\cdots\otimes\Omega_n^\lambda)(\rho_{L_1}^\lambda\otimes \rho_{L_2}^\lambda\otimes\cdots\otimes\rho_{L_r}^\lambda),\end{equation}
where $\{p_\Lambda(\lambda)\}$ is a probability distribution, $\Omega_j^\lambda$ is a quantum channel of the subsystem $A_j$ \cite{NWRP}.
So the number $n$ of nodes corresponds to the number of
components in the composite system involved in the network, the number $r$ of factors contained in the product state  corresponds to the number of sources. Denote by $\sharp(L_i)$ the number  of nodes involved in the source $\rho_i$.  The most shared nodes $k=\max \limits_i\sharp(L_i)$ is called the {\it depth of the network}, which reflects the fact that  the quantum network state with respect to this network  can be obtained  by preparing the states in at most $k$-partite system, and the corresponding network states of the form in Eq.(6) are called  network states with depth $k$.

\begin{example} Consider the scenario that we have three networks, net 1, net 2  and net 3, as in Fig.1, and three 6-qubit states,
 $$\rho_a = |\psi_a\rangle\langle\psi_a|, \ \ \rho_b = |\psi_b\rangle\langle\psi_b|,\ \  \rho_c = |\psi_c\rangle\langle\psi_c|$$
determined by these networks, where $$\begin{array}{rl}|\psi_a\rangle=&\frac{1}{8}(|000000\rangle+|000011\rangle+|001100\rangle+|001111\rangle\\
	&+|110000\rangle+|110011\rangle+|111100\rangle+|111111\rangle),\end{array}$$
$$|\psi_b\rangle=\frac{1}{4}(|000000\rangle+|000111\rangle+|111000\rangle+|111111\rangle),$$ $$|\psi_c\rangle=\frac{1}{4}(|000000\rangle+|000011\rangle+|111100\rangle+|111111\rangle).$$
The task is that, for each $j\in\{1,2,3\}$, find $x\in\{a,b,c\}$ such that $\rho_x$ is exactly the network state corresponding to the network net $j$.
\end{example}

\begin{widetext}
	\begin{figure}[ht]
  \centering
   \begin{minipage}{\textwidth}
  \includegraphics[width=14cm]{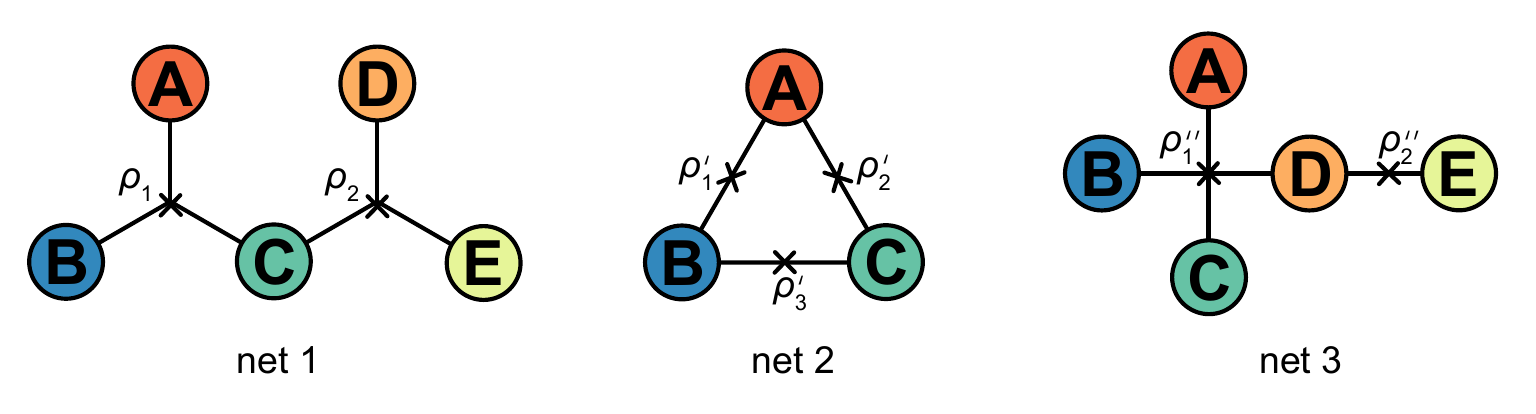}\\
  \caption{Three different quantum networks.}\label{net}
   \end{minipage}
\end{figure}
\end{widetext}

In this example, we also take $s = -\infty$ for an illustration. To do this, we observe that,  for the network net 1 in Fig.1,
parties $A, B$ and $C$ share a source state $\rho_{1}$, while parties $C, D$ and $E$ share a source state $\rho_{2}$. This indicates that the depth of the network net 1 is 3 and the corresponding network state must be 3-producible.  Similarly, the depth of the network net 2 is 2 and the corresponding network state  is 2-producible, while   the depth of the network net 3 is 4 and the corresponding network state is 4-producible.

It is known by Theorem 4, for a  state $\rho$ in a $6$-partite system and $1\leq k\leq 5$, $\rho$ is $k$-producible implies that $$\sum_{b=1}^{3} \sum_{n=1}^{2} I^s(\rho, {\bf P}_{6,n}^{(b)})\leq I_k,$$
where
\begin{widetext}
	$$I_k=\begin{cases}
	6k(\kappa - \frac{1}{2}) + 6(2\kappa-1) & {\rm if}\ \ 6=pk; \\
	(36+p^2k^2+pk^2 -12pk) (\kappa - \frac{1}{2}) + 6(2\kappa-1) & {\rm if}\ \ 0<6-pk<k,
\end{cases}
$$
 $p=[\frac{6}{k}]$.
\end{widetext}
The concrete values of $I_k$ for each $k$ is list in Table IX.
\begin{table}[ht]
    \centering
    \setlength{\tabcolsep}{12pt}
    \renewcommand{\arraystretch}{1.5}
    \begin{tabular}{l|ccccc}
        \hline
        $k$ & 1 & 2 & 3 & 4 & 5 \\
        \hline
        $I_k$ & 9 & 12 & 15 & 16 & 19 \\
        \hline
    \end{tabular}
    \caption{Threshold values $I_k$ for $k$-producible states in 6-partite quantum systems.}
    \label{network}
\end{table}

Regard $\rho_a$, $\rho_b$ and $\rho_c$ as $6$-partite states.  A direct calculation gives
$$
I_a = \sum_{b=1}^{3} \sum_{n=1}^{2} I^{-\infty }(\rho_a, {\bf P}_{6,n}^{(b)})=12,$$
$$I_b = \sum_{b=1}^{3} \sum_{n=1}^{2} I^{-\infty }(\rho_b, {\bf P}_{6,n}^{(b)})=15$$
and $$I_c = \sum_{b=1}^{3} \sum_{n=1}^{2} I^{-\infty }(\rho_c, {\bf P}_{6,n}^{(b)})=16.
$$
Comparing with Table IX, we have
 $$I_1<I_a \leq  I_k \quad\mbox{\rm for } k\leq 2;$$
 $$I_1<I_2<  I_b \leq I_k\quad\mbox{\rm for }  k\geq 3$$
 and
 $$ I_1<I_2<I_3< I_c \leq I_k \quad\mbox{\rm for }  k\geq 4.$$
Then, by Theorem 4, it is easily seen  that
$\rho_c$ is not 3-producible and thus is the network state corresponding to the network of depth greater than 3. Consequently, $\rho_c$ is the state corresponding to net 3. Similarly, one concludes that  $\rho_b $ is the network state corresponding to the network net 1 and $\rho_a$ is corresponding to net 2. This completes the given task.
\hfill$\Box$

\section{Conclusion}

We establish two criteria based on MUMs and the generalized Wigner-Yanase skew information to detect $k$-nonseparability and $k$-partite entanglement, respectively, in multipartite quantum systems. Through two examples, we found that when the parameter takes the limiting value $s = -\infty$, the generalized Wigner-Yanase skew information exhibits the best detection performance, being more effective in detecting multipartite quantum entanglement compared to other values of $s$. These findings could offer new perspectives on detecting multipartite quantum entanglement and may also be applicable to the detection of other multipartite quantum correlations.
 It is also worth noting that the upper bounds of the inequalities obtained in our main theorems are independent of the parameter $s$ of the generalized Wigner-Yanase skew information $I^s(\rho,A)$. This is because we used the inequality  $I^s(\rho,A)\leq V(\rho, A)$ during the proof process. However, given the superior performance at the limiting case $s = -\infty$, an important and challenging direction for future research is whether one can establish tighter upper bounds that depend on $s$, thereby further enhancing the detection capability of multipartite entanglement.

\section{Acknowledgments}

The authors wish to give their thanks to the referees for many helpful comments to improve the original paper. This work is partially supported by the National Natural
Science Foundation of China (12571338, 12171290, 12071336).

\end{document}